\documentclass[11pt,a4paper]{article}
\usepackage[latin9]{inputenc}
\usepackage{xcolor}
\usepackage{booktabs}
\usepackage{amsmath}
\usepackage{graphicx}

\makeatletter

\pdfpageheight\paperheight
\pdfpagewidth\paperwidth

\providecommand{\tabularnewline}{\\}

\usepackage{pos}
\usepackage[compat=1.1.0]{tikz-feynman}

\title{Flavour hierarchies and $B$-anomalies in a twin Pati-Salam theory of flavour}
\ShortTitle{Twin Pati-Salam}

\author*[\dag]{Mario Fernández Navarro}
\notes{\note{The author expresses sincere 
thanks to the organisers, in particular to Syuhei Iguro, for the invitation and their hospitality and 
acknowledges the European Union's Horizon 2020 Research and Innovation programme under Marie Sk\l{}odowska-Curie grant agreement HIDDeN European ITN project (H2020-MSCA-ITN-2019//860881-HIDDeN). 
The author thanks Xavier Ponce Díaz for carefully reading this manuscript.}}

\affiliation{School of Physics \& Astronomy, University of Southampton\\
 Southampton SO17 1BJ, UK}

\emailAdd{M.F.Navarro@soton.ac.uk}

\abstract{In this proceedings, based on \href{https://arxiv.org/abs/2209.00276}{arXiv:2209.00276}, I present a model which can simultaneously explain and connect the flavour hierarchies of the Standard Model with flavour anomalies in $B$-physics. 
I will briefly introduce the model and highlight the main features, including a common origin of Yukawa couplings and vector leptoquark $U_{1}$ couplings to Standard Model fermions. 
A GIM-like mechanism allows for large leptoquark couplings which can explain the $B$-anomalies, while protecting from the most dangerous FCNCs that could be mediated by a heavy coloron and $Z'$. 
Finally, I will highlight some of the most promising signals at low energy processes which can test the model in the upcoming future. The analysis has been updated with the late 2022 measurements of $R_{K^{(*)}}$ by LHCb.}

\FullConference{%
  8th Symposium on Prospects in the Physics of Discrete Symmetries (DISCRETE 2022)\\
  7-11 November, 2022\\
  Baden-Baden, Germany
}

\@ifundefined{showcaptionsetup}{}{%
 \PassOptionsToPackage{caption=false}{subfig}}
\usepackage{subfig}
\makeatother

\begin{document}
\setcounter{page}{0}
\maketitle

\section{Introduction}

Fundamental fermions in the Standard Model (SM) come in three copies,
denoted as ``flavours'', which share universal gauge interactions
but have different masses and mixings, also known as flavour parameters.
The origin of flavour in the SM remains as a complete mystery, as
it lacks of any dynamical explanation to the high number of flavour
parameters and their hierarchical patterns. A theory of flavour beyond
the SM should provide a solution to the long-lasting ``flavour puzzle''.

Simultaneously, the non-universal structure of such a theory of flavour
could leave its imprints in flavour physics observables, which are
becoming accessible up to a high precision level in the current generation
of colliders and meson factories. In this direction, a conspicuous
series of anomalies in flavour observables emerged in the last years.
Of particular interest were the anomalies in $B$-meson decays, namely
the $R_{K^{(*)}}$ \cite{LHCb:2021trn,LHCb:2017avl} and $R_{D^{(*)}}$
\cite{Belle:2019gij} ratios hinting for the breaking of the SM lepton
flavour universality (LFU), a feature which might well be a low energy
signal of the theory addressing the origin of flavour hierarchies.
The origin of flavour was connected to the $B$-anomalies in models
featuring $Z'$ bosons \cite{King:2018fcg,Navarro:2021sfb} or scalar
leptoquarks \cite{deMedeirosVarzielas:2018bcy}.

Interestingly, the vector leptoquark $U_{1}(\boldsymbol{3},\boldsymbol{1},2/3)$
was the only leptoquark capable of addressing both ``$B$-anomalies''
simultaneously \cite{Angelescu:2021lln} (and further improve the
global fit to $b\rightarrow s\ell\ell$ data via loop effects \cite{Crivellin:2018yvo}).
Moreover, hierarchical couplings of $U_{1}$ to SM fermions were required
to simultaneously address both anomalies, which might well be related
to the flavour hierarchies of the SM in the UV theory. The model presented
here was built to address this 2021 picture of $B$-anomalies via
$U_{1}$ exchange, and connect their origin to the origin of Yukawa
couplings in the SM. However, in late 2022 the LHCb collaboration
updated their measurements of $R_{K^{(*)}}$ \cite{LHCb:2022qnv},
which are now in good agreement with the SM predictions. We consider
these new measurements in this analysis, and show that the model is
still compatible with the new data.

The massive $U_{1}$ leptoquark is well known to arise from spontaneous
breaking of the traditional Pati-Salam gauge group (PS). However,
this breaking happening at the TeV scale is at odds with current bounds
from $K_{L}\rightarrow\mu e$ and other processes, unless further
model building is performed (see e.g.~\cite{Calibbi:2017qbu,Blanke:2018sro}
for alternative PS embeddings). In this direction, a 4321 gauge symmetry,

\begin{equation}
G_{4321}\equiv SU(4)\times SU(3)_{c}^{'}\times SU(2)_{L}\times U(1)_{Y'}\,,
\end{equation}
was proposed as a possible origin of $U_{1}$ at the TeV scale \cite{DiLuzio:2017vat,DiLuzio:2018zxy,Cornella:2019hct,Cornella:2021sby}.
UV completions of the 4321 group had been proposed in recent years
to connect the flavour hierarchies of the SM with the $B$-anomalies
via $U_{1}$ \cite{Bordone:2017bld,Fuentes-Martin:2022xnb}, however
all of them consider that the third family of fermions transforms
under the TeV scale $SU(4)$ (the others are singlets), which leads
to strong constraints via high-$p_{T}$ production of heavy gauge
bosons arising after 4321 breaking \cite{Baker:2019sli,Aebischer:2022oqe}. 

Instead, we proposed a twin Pati-Salam theory of flavour \cite{King:2021jeo,FernandezNavarro:2022gst}
which features a TeV scale 4321 group under which all SM fermions
are treated in the same way, they are singlets under the TeV scale
$SU(4)$. Three families of vector-like fermions are charged under
$SU(4)$ and couple to $U_{1}$, generating effective couplings to
SM fermions via mixing. Simultaneously, the choice of the scalar sector
and the twin PS symmetry forbid SM-like Yukawa couplings, which are
generated via the same mixing that led to the leptoquark couplings.
This provides a link between $B$-physics and the theory of flavour.
In this manuscript, we will briefly introduce the model and summarise
the main phenomenological consequences.

\section{The model}

The theory is based on two copies of the traditional PS group,
\begin{table}[t]
\begin{centering}
\begin{tabular}{lccccccc}
\toprule 
Field & $SU(4)_{PS}^{I}$ & $SU(2)_{L}^{I}$ & $SU(2)_{R}^{I}$ & $SU(4)_{PS}^{II}$ & $SU(2)_{L}^{II}$ & $SU(2)_{R}^{II}$ & $Z_{4}$\tabularnewline
\midrule
\midrule 
$\psi_{1,2,3}$ & $\mathbf{1}$ & $\mathbf{1}$ & $\mathbf{1}$ & $\mathbf{4}$ & $\mathbf{2}$ & $\mathbf{1}$ & $\alpha$, 1, 1\tabularnewline
$\psi_{1,2,3}^{c}$ & $\mathbf{1}$ & $\mathbf{1}$ & $\mathbf{1}$ & $\mathbf{\overline{4}}$ & $\mathbf{1}$ & $\mathbf{\overline{2}}$ & $\alpha$, $\alpha^{2}$, 1\tabularnewline
\midrule 
$\psi_{4,5,6}$ & $\mathbf{4}$ & $\mathbf{2}$ & $\mathbf{1}$ & $\mathbf{1}$ & $\mathbf{1}$ & $\mathbf{1}$ & 1, 1, $\alpha$\tabularnewline
$\overline{\psi}_{4,5,6}$ & $\mathbf{\overline{4}}$ & $\mathbf{\overline{2}}$ & $\mathbf{1}$ & $\mathbf{1}$ & $\mathbf{1}$ & $\mathbf{1}$ & 1, 1, $\alpha^{3}$\tabularnewline
$\psi_{4,5,6}^{c}$ & $\mathbf{\overline{4}}$ & $\mathbf{1}$ & $\mathbf{\overline{2}}$ & $\mathbf{1}$ & $\mathbf{1}$ & $\mathbf{1}$ & 1, 1, $\alpha$\tabularnewline
$\overline{\psi^{c}}_{4,5,6}$ & $\mathbf{4}$ & $\mathbf{1}$ & $\mathbf{2}$ & $\mathbf{1}$ & $\mathbf{1}$ & $\mathbf{1}$ & 1, 1, $\alpha^{3}$\tabularnewline
\midrule 
$\phi$ & $\mathbf{4}$ & $\mathbf{2}$ & $\mathbf{1}$ & $\mathbf{\overline{4}}$ & $\mathbf{\overline{2}}$ & $\mathbf{1}$ & 1\tabularnewline
$\overline{\phi}$,$\,\overline{\phi'}$ & $\mathbf{\overline{4}}$ & $\mathbf{1}$ & $\mathbf{\overline{2}}$ & $\mathbf{4}$ & $\mathbf{1}$ & $\mathbf{2}$ & 1, $\alpha^{2}$\tabularnewline
\midrule
$H$ & $\mathbf{\overline{4}}$ & $\mathbf{\overline{2}}$ & $\mathbf{1}$ & $\mathbf{4}$ & $\mathbf{1}$ & $\mathbf{2}$ & 1\tabularnewline
$\overline{H}$ & $\mathbf{4}$ & $\mathbf{1}$ & $\mathbf{2}$ & $\mathbf{\overline{4}}$ & $\mathbf{\overline{2}}$ & $\mathbf{1}$ & 1\tabularnewline
\midrule
$\Omega_{15}$ & $\mathbf{15}$ & $\mathbf{1}$ & $\mathbf{1}$ & $\mathbf{1}$ & $\mathbf{1}$ & $\mathbf{1}$ & 1\tabularnewline
\bottomrule
\end{tabular}
\par\end{centering}
\caption{The field content under $G_{422}^{I}\times G_{422}^{II}\times Z_{4}$.
The model consists of three left-handed chiral fermion families $\psi_{1,2,3}$,
$\psi_{1,2,3}^{c}$ under the second PS group, plus three VL fermion
families $\psi_{4,5,6}$, $\psi_{4,5,6}^{c}$ and their conjugates
under the first PS group. Personal Higgs doublets are contained in
$H$, $\bar{H}$, one for each second and third family charged fermion.
The Higgs singlets in $\phi$, $\bar{\phi}$ are called Yukons. In
addition to the fields shown here, we require further high energy
Higgs fields (not shown) whose VEVs will break the second PS group
at a high scale, leaving the first unbroken (see \cite{King:2021jeo,FernandezNavarro:2022gst}).
We also need further Higgs fields, which break the two left-right
gauge groups into their diagonal subgroup.\label{tab:The-field-content_ExtendedModel}}

\end{table}

\begin{equation}
G_{422}=SU(4)\times SU(2)_{L}\times SU(2)_{R}\,,
\end{equation}
plus a shaping discrete symmetry $Z_{4}$, i.e.

\begin{equation}
G_{422}^{I}\times G_{422}^{II}\times Z_{4}\,.
\end{equation}
The content of the model is depicted in Table~\ref{tab:The-field-content_ExtendedModel}.
We assume that three chiral fermion families transform under the second
PS group, $G_{422}^{II}$, which is broken at a high scale $M_{\mathrm{High}}\gtrsim1\,\mathrm{PeV}$
down to $G_{4321}$, the latter bound imposed by $K_{L}\rightarrow\mu e$.
On the other hand, three vector-like (VL) fermion families transform
under the first PS group, $G_{422}^{I}$, in such a way that they
couple to the TeV scale leptoquark $U_{1}$ arising after the breaking
of $G_{4321}$ down to the SM. They further couple to a TeV scale
coloron $g'(\mathbf{8},\mathbf{1},0)$ and $Z'(\mathbf{1},\mathbf{1},0)$
that also arise after the breaking of $G_{4321}$. The renormalizable
mass Lagrangian is given by 
\begin{align}
\begin{aligned}\mathcal{L}_{\mathrm{mass}}^{ren} & =y_{ia}^{\psi}\overline{H}\psi_{i}\psi_{a}^{c}+y_{a3}^{\psi}H\psi_{a}\psi_{3}^{c}+x_{ia}^{\psi}\phi\psi_{i}\overline{\psi}_{a}+x_{a2}^{\psi^{c}}\overline{\psi_{a}^{c}\phi'}\psi_{2}^{c}+x_{a3}^{\psi^{c}}\overline{\psi_{a}^{c}\phi}\psi_{3}^{c}+x_{16}^{\psi}\phi\psi_{1}\overline{\psi}_{6}+x_{61}^{\psi^{c}}\overline{\psi_{6}^{c}\phi}\psi_{1}^{c} & {}\\
{} & +M_{ab}^{\psi}\psi_{a}\overline{\psi_{b}}+M_{ab}^{\psi^{c}}\psi_{a}^{c}\overline{\psi_{b}^{c}}+M_{66}^{\psi}\psi_{6}\overline{\psi_{6}}+M_{66}^{\psi^{c}}\psi_{6}^{c}\overline{\psi_{6}^{c}} & {}\\
{} & +\lambda_{15}^{aa}\Omega_{15}\psi_{a}\overline{\psi_{a}}+\lambda_{15}^{66}\Omega_{15}\psi_{6}\overline{\psi_{6}}+\bar{\lambda}_{15}^{aa}\Omega_{15}\psi_{a}^{c}\overline{\psi_{a}^{c}}+\bar{\lambda}_{15}^{66}\Omega_{15}\psi_{6}^{c}\overline{\psi_{6}^{c}}+\mathrm{h.c.}\,, & {}
\end{aligned}
\label{eq: full_lagrangian}
\end{align}
where $i,j=2,3$ and $a,b=4,5$. It can be seen that due to the choice
of the scalar sector and the twin PS symmetry, SM-like Yukawa couplings
are not present. The Lagrangian above can be written as a mass matrix,
and its diagonalisation will lead to effective Yukawa couplings for
SM fermions.

We shall refer to the Higgs doublets contained in $H$, $\bar{H}$
as personal Higgs doublets, since under the SM decomposition there
will be a separate Higgs for each fermion mass as we shall see shortly.
The Higgs singlet fields in $\phi$ , $\bar{\phi}$ are called Yukons,
since they are necessary to generate the effective Yukawa couplings.
They develop different VEVs in their quark and lepton components ($\phi_{3}$
and $\phi_{1}$), breaking $G_{4321}$ at the TeV scale down to the
SM.

\subsection{Effective Yukawa couplings and leptoquark couplings}

The diagonalisation of the full mass matrix leads to mixing between
VL and chiral fermions. Each VL family mixes with one chiral family,
with mixing angles given by

\begin{figure}
\begin{raggedright}
\subfloat[]{\begin{centering}
\resizebox{0.47\textwidth}{!}{
\begin{tikzpicture}
	\begin{feynman}
		\vertex (a) {\(Q_{3}\)};
		\vertex [right=18mm of a] (b);
		\vertex [right=of b] (c) [label={ [xshift=0.1cm, yshift=0.1cm] \small $M^{Q}_{4}$}];
		\vertex [right=of c] (d);
		\vertex [right=of d] (e) {\(u^{c}_{3}\)};
		\vertex [above=of b] (f1) {\(\phi_{3}\)};
		\vertex [above=of d] (f2) {\(H_{t}\)};
		\diagram* {
			(a) -- [fermion] (b) -- [charged scalar] (f1),
			(b) -- [edge label'=\(\overline{Q_{4}}\)] (c),
			(c) -- [edge label'=\(Q_{4}\), inner sep=6pt, insertion=0] (d) -- [charged scalar] (f2),
			(d) -- [fermion] (e),
	};
	\end{feynman}
\end{tikzpicture}
}
\par\end{centering}
}$\quad$\subfloat[]{\begin{centering}
\resizebox{0.47\textwidth}{!}{
\begin{tikzpicture}
	\begin{feynman}
		\vertex (a) {\(Q_{3}\)};
		\vertex [right=13mm of a] (b);
		\vertex [right=11mm of b] (c) [label={ [xshift=0.1cm, yshift=0.1cm] \small $M^{Q}_{4}$}];
		\vertex [right=11mm of c] (d);
		\vertex [right=11mm of d] (e) [label={ [xshift=0.1cm, yshift=0.1cm] \small $M^{L}_{4}$}];
		\vertex [right=11mm of e] (f);
		\vertex [right=11mm of f] (g) {\(L_{3}\)};
		\vertex [above=14mm of b] (f1) {\(\phi_{3}\)};
		\vertex [above=14mm of d] (f2) {\(U_{1}\)};
		\vertex [above=14mm of f] (f3) {\(\phi_{1}\)};
		\diagram* {
			(a) -- [fermion] (b) -- [charged scalar] (f1),
			(b) -- [edge label'=\(\overline{Q}_{4}\)] (c),
			(c) -- [edge label'=\(Q_{4}\), inner sep=6pt, insertion=0] (d) -- [boson, red] (f2),
			(d) -- [edge label'=\(L_{4}\), inner sep=6pt] (e),
			(e) -- [edge label'=\(\overline{L}_{4}\), insertion=0] (f) -- [charged scalar] (f3),
			(f) -- [fermion] (g),
	};
	\end{feynman}
\end{tikzpicture}
}
\par\end{centering}
}
\par\end{raggedright}
\begin{raggedright}
{\scriptsize{}}\subfloat[\label{fig:2-3_mixing}]{\begin{centering}
\resizebox{0.47\textwidth}{!}{
\begin{tikzpicture}
	\begin{feynman}
		\vertex (a) {\(Q_{i}\)};
		\vertex [right=18mm of a] (b);
		\vertex [right=of b] (c) [label={ [xshift=0.1cm, yshift=0.1cm] \small $M^{\psi^{c}}_{4}$}];
		\vertex [right=of c] (d);
		\vertex [right=of d] (e) {\(u^{c}_{j}\)};
		\vertex [above=of b] (f1) {\(H_{c}\)};
		\vertex [above=of d] (f2) {\(\overline{\phi}_{3}\)};
		\diagram* {
			(a) -- [fermion] (b) -- [charged scalar] (f1),
			(b) -- [edge label'=\(\overline{u^{c}_{4}}\)] (c),
			(c) -- [edge label'=\(u^{c}_{4}\), inner sep=6pt, insertion=0] (d) -- [charged scalar] (f2),
			(d) -- [fermion] (e),
	};
	\end{feynman}
\end{tikzpicture}
}
\par\end{centering}
{\scriptsize{}}{\scriptsize\par}}{\scriptsize{}$\quad$}\subfloat[]{\begin{centering}
\resizebox{0.47\textwidth}{!}{
\begin{tikzpicture}
	\begin{feynman}
		\vertex (a) {\(u^{c}_{3}\)};
		\vertex [right=13mm of a] (b);
		\vertex [right=11mm of b] (c) [label={ [xshift=0.1cm, yshift=0.1cm] \small $M^{\psi^{c}}_{4}$}];
		\vertex [right=11mm of c] (d);
		\vertex [right=11mm of d] (e) [label={ [xshift=0.1cm, yshift=0.1cm] \small $M^{\psi^{c}}_{4}$}];
		\vertex [right=11mm of e] (f);
		\vertex [right=11mm of f] (g) {\(e^{c}_{3}\)};
		\vertex [above=14mm of b] (f1) {\(\overline{\phi}_{3}\)};
		\vertex [above=14mm of d] (f2) {\(U_{1}\)};
		\vertex [above=14mm of f] (f3) {\(\overline{\phi}_{1}\)};
		\diagram* {
			(a) -- [fermion] (b) -- [charged scalar] (f1),
			(b) -- [edge label'=\(\overline{u^{c}_{4}}\)] (c),
			(c) -- [edge label'=\(u^{c}_{4}\), inner sep=6pt, insertion=0] (d) -- [boson, red] (f2),
			(d) -- [edge label'=\(e^{c}_{4}\), inner sep=6pt] (e),
			(e) -- [edge label'=\(\overline{e^{c}_{4}}\), insertion=0] (f) -- [charged scalar] (f3),
			(f) -- [fermion] (g),
	};
	\end{feynman}
\end{tikzpicture}
}
\par\end{centering}
}
\par\end{raggedright}
\caption{\textbf{\textit{(Top)}} The diagram on the left leads to the mass
of the top quark, and similar diagrams can be written for all third
family charged fermions. The diagram on the right leads to effective
couplings of the third family left-handed fermions to the $U_{1}$
vector leptoquark. \textbf{\textit{(Bottom) }}The diagram on the left
leads to the mass of second family charged fermions, $i,j=2,3$, and
their mixing with the third family. The diagram on the right leads
to effective couplings of the third family right-handed fermions to
the $U_{1}$ vector leptoquark, which are suppressed as they are connected
to the origin of second family fermion masses.\label{fig:Effective_Yukawas_U1}}

\end{figure}
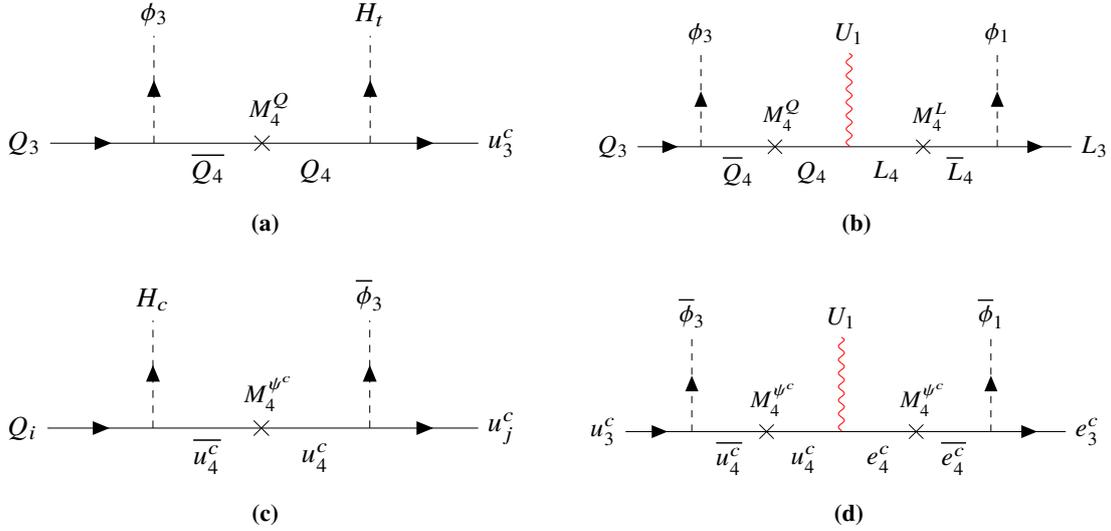

{\footnotesize{}
\begin{equation}
s_{34}^{Q,L}=\frac{x_{34}^{\psi}\left\langle \phi_{3,1}\right\rangle }{\sqrt{\left(x_{34}^{\psi}\left\langle \phi_{3,1}\right\rangle \right)^{2}+\left(M_{4}^{Q,L}\right)^{2}}}\,,\:s_{25}^{Q,L}=\frac{x_{25}^{\psi}\left\langle \phi_{3,1}\right\rangle }{\sqrt{\left(x_{25}^{\psi}\left\langle \phi_{3,1}\right\rangle \right)^{2}+\left(M_{5}^{Q,L}\right)^{2}}}\,,\:s_{16}^{Q,L}=\frac{x_{16}^{\psi}\left\langle \phi_{3,1}\right\rangle }{\sqrt{\left(x_{16}^{\psi}\left\langle \phi_{3,1}\right\rangle \right)^{2}+\left(M_{6}^{Q,L}\right)^{2}}}\,,\label{eq:VL_chiral_mixing}
\end{equation}
}which further lead to effective Yukawa couplings and $U_{1}$ couplings
for SM fermions, as depicted in the diagrams of Fig.~\ref{fig:Effective_Yukawas_U1}.
It can be seen that both the Yukawa couplings and $U_{1}$ couplings
originate via the same physics, and are connected via the same mixing
angles.

The large top mass requires $\left\langle \phi_{3}\right\rangle /M_{4}^{Q}\sim1$,
hence the 3-4 mixing angle is expected to be very large. This is in
good agreement with the expectation of $R_{D^{(*)}}$, which requires
large couplings of $U_{1}$ to third family fermions, featuring a
nice connection between the flavour puzzle and $B$-anomalies only
present in our model. Moreover, the origin of second family masses
requires heavy masses for EW singlets VL fermions, leading to the
hierarchy of VL masses,

\begin{equation}
M_{a}^{\psi}\ll M_{a}^{\psi^{c}}
\end{equation}
for $a=4,5,6$. This way, leptoquark currents involving right-handed
(EW singlets $\psi^{c}$) chiral fermions are suppressed by small
mixing angles, connected to the origin of second family fermion masses,
leading to a model which dominantly predicts left-handed leptoquark
currents as preferred by the global fits of $B$-anomalies (see e.g.~\cite{Geng:2021nhg,Angelescu:2021lln,Altmannshofer:2021qrr}).
This way, we obtain another nice connection between the flavour puzzle
and $B$-anomalies, see the bottom diagrams in Fig.~\ref{fig:Effective_Yukawas_U1}.

\subsection{GIM-like mechanism and FCNCs}

The explanation of $R_{D^{(*)}}$ requires large flavour-violating
$U_{1}$ couplings that compete with the tree-level SM charged current.
Obtaining such large couplings while suppressing possible FCNCs mediated
by the $Z'$ and $g'$ is a challenge for 4321 models, which is usually
addressed via adopting an ad-hoc flavour structure, calling for an
understanding in terms of a UV theory which the twin PS model can
provide.
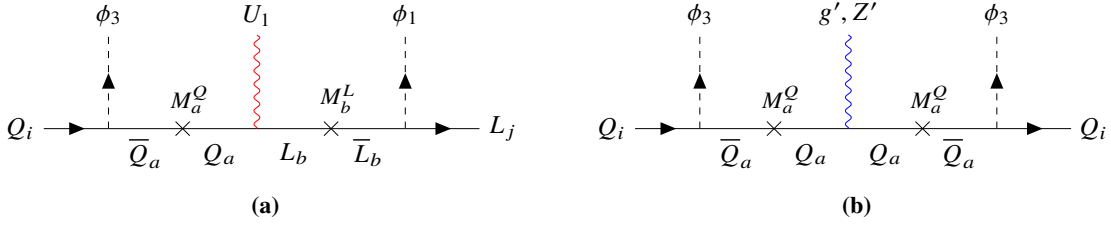
\begin{figure}
\begin{raggedright}
\subfloat[]{\begin{centering}
\resizebox{0.47\textwidth}{!}{
\begin{tikzpicture}
	\begin{feynman}
		\vertex (a) {\(Q_{i}\)};
		\vertex [right=13mm of a] (b);
		\vertex [right=11mm of b] (c) [label={ [xshift=0.1cm, yshift=0.1cm] \small $M^{Q}_{a}$}];
		\vertex [right=11mm of c] (d);
		\vertex [right=11mm of d] (e) [label={ [xshift=0.1cm, yshift=0.1cm] \small $M^{L}_{b}$}];
		\vertex [right=11mm of e] (f);
		\vertex [right=11mm of f] (g) {\(L_{j}\)};
		\vertex [above=14mm of b] (f1) {\(\phi_{3}\)};
		\vertex [above=14mm of d] (f2) {\(U_{1}\)};
		\vertex [above=14mm of f] (f3) {\(\phi_{1}\)};
		\diagram* {
			(a) -- [fermion] (b) -- [charged scalar] (f1),
			(b) -- [edge label'=\(\overline{Q}_{a}\)] (c),
			(c) -- [edge label'=\(Q_{a}\), inner sep=6pt, insertion=0] (d) -- [boson, red] (f2),
			(d) -- [edge label'=\(L_{b}\), inner sep=6pt] (e),
			(e) -- [edge label'=\(\overline{L}_{b}\), insertion=0] (f) -- [charged scalar] (f3),
			(f) -- [fermion] (g),
	};
	\end{feynman}
\end{tikzpicture}
}
\par\end{centering}
}$\quad$\subfloat[\label{fig:couplings_g'_Z'}]{\begin{centering}
\resizebox{0.47\textwidth}{!}{
\begin{tikzpicture}
	\begin{feynman}
		\vertex (a) {\(Q_{i}\)};
		\vertex [right=13mm of a] (b);
		\vertex [right=11mm of b] (c) [label={ [xshift=0.1cm, yshift=0.1cm] \small $M^{Q}_{a}$}];
		\vertex [right=11mm of c] (d);
		\vertex [right=11mm of d] (e) [label={ [xshift=0.1cm, yshift=0.1cm] \small $M^{Q}_{a}$}];
		\vertex [right=11mm of e] (f);
		\vertex [right=11mm of f] (g) {\(Q_{i}\)};
		\vertex [above=14mm of b] (f1) {\(\phi_{3}\)};
		\vertex [above=14mm of d] (f2) {\(g',Z'\)};
		\vertex [above=14mm of f] (f3) {\(\phi_{3}\)};
		\diagram* {
			(a) -- [fermion] (b) -- [charged scalar] (f1),
			(b) -- [edge label'=\(\overline{Q}_{a}\)] (c),
			(c) -- [edge label'=\(Q_{a}\), inner sep=6pt, insertion=0] (d) -- [boson, blue] (f2),
			(d) -- [edge label'=\(Q_{a}\), inner sep=6pt] (e),
			(e) -- [edge label'=\(\overline{Q}_{a}\), insertion=0] (f) -- [charged scalar] (f3),
			(f) -- [fermion] (g),
	};
	\end{feynman}
\end{tikzpicture}
}
\par\end{centering}
}
\par\end{raggedright}
\caption{\textbf{\textit{(Left)}} Flavour off-diagonal effective couplings
of the $U_{1}$ vector leptoquark to SM fermions. \textbf{\textit{(Right)}}\textit{
}Flavour diagonal effective couplings of the heavy coloron and $Z'$
to SM quarks. Similar diagrams are obtained for leptons with the $Z'$
boson.\label{fig:FCCC_FCNC}}

\end{figure}

The scalar $\Omega_{15}$ transforming in the adjoint of $SU(4)^{I}$
splits the bare masses of VL quarks and leptons $M_{aa}^{Q,L}$, which
otherwise would be identical due to the twin PS symmetry. The mass
matrix of VL fermions is diagonalised via different transformations
for quark and leptons, leading to large flavour-violating $U_{1}$
couplings,{\footnotesize{}
\begin{equation}
\left.\begin{array}{c}
\left(\begin{array}{ccc}
M_{4}^{Q} & 0 & 0\\
0 & M_{5}^{Q} & 0\\
0 & 0 & M_{6}^{Q}
\end{array}\right)=V_{45}^{Q}\left(\begin{array}{ccc}
M_{44}^{Q} & M_{45}^{\psi} & 0\\
M_{54}^{\psi} & M_{55}^{Q} & 0\\
0 & 0 & M_{66}^{Q}
\end{array}\right)V_{45}^{\bar{Q}\dagger}\\
\left(\begin{array}{ccc}
M_{4}^{L} & 0 & 0\\
0 & M_{5}^{L} & 0\\
0 & 0 & M_{6}^{L}
\end{array}\right)=V_{45}^{L}\left(\begin{array}{ccc}
M_{44}^{L} & M_{45}^{\psi} & 0\\
M_{54}^{\psi} & M_{55}^{L} & 0\\
0 & 0 & M_{66}^{L}
\end{array}\right)V_{45}^{\bar{L}\dagger}
\end{array}\right\} \Rightarrow\frac{g_{4}}{\sqrt{2}}Q_{a}^{\dagger}\gamma_{\mu}\left(\begin{array}{ccc}
c_{\theta_{LQ}} & -s_{\theta_{LQ}} & 0\\
s_{\theta_{LQ}} & c_{\theta_{LQ}} & 0\\
0 & 0 & 1
\end{array}\right)L{}_{b}U_{1}^{\mu}+\mathrm{h.c.}
\end{equation}
}where the zeroes are enforced by the $Z_{4}$ symmetry. This way,
we obtain a CKM-like matrix $W_{LQ}\equiv V_{45}^{Q}V_{45}^{L\dagger}$
in $SU(4)^{I}$ flavour space. Remarkably, this way we obtain large
flavour-violating leptoquark currents in VL quark-lepton flavour space,
while neutral currents mediated by $g'$ and $Z'$ remain flavour
universal. Afterwards, 3-4, 2-5 and 1-6 mixing is performed (see Eq.~\eqref{eq:VL_chiral_mixing}),
inducing the required structure of $U_{1}$ couplings that contribute
to LFU ratios. On the other hand, neutral currents become flavour
diagonal, controlled by the mixing angles $s_{34}^{Q,L}$, $s_{25}^{Q,L}$
and $s_{16}^{Q,L}$, see Fig.~\ref{fig:FCCC_FCNC}.
\begin{itemize}
\item If $s_{25}^{Q,L}\approx s_{16}^{Q,L}$ then there is no 1-2 tree-level
FCNCs, featuring a GIM-like mechanism which allows for large flavour-violating
leptoquark couplings to explain the anomalies but protects from the
most dangerous FCNCs. A down-aligned flavour structure is achieved
in the 2-3 sector (see \cite{FernandezNavarro:2022gst}), protecting
from tree-level contributions to $B_{s}-\bar{B}_{s}$ meson mixing.
\item $s_{34}^{Q}\approx1$ is motivated by the top mass and $R_{D^{(*)}}$,
as anticipated before. $s_{16}^{Q}\approx0.2$ is compatible with
bounds from coloron production at LHC.
\item We choose a suitable benchmark $\left\langle \phi_{3,1}\right\rangle \approx0.6\,\mathrm{TeV},\,0.3\,\mathrm{TeV}$
and $M_{44}^{Q,L}<M_{55}^{Q,L},M_{66}^{Q,L}\approx1.2\,\mathrm{TeV}-\,0.8\,\mathrm{TeV}$\textcolor{orange}{{}
}and we explore the parameter space of $x_{34}^{\psi}$ and $x_{25}^{\psi}$
(see \cite{FernandezNavarro:2022gst} for further details about the
benchmark).
\end{itemize}

\section{Phenomenology}

\begin{figure}
\subfloat[\label{fig:Bs_Mixing_ParameterSpace}]{\includegraphics[scale=0.37]{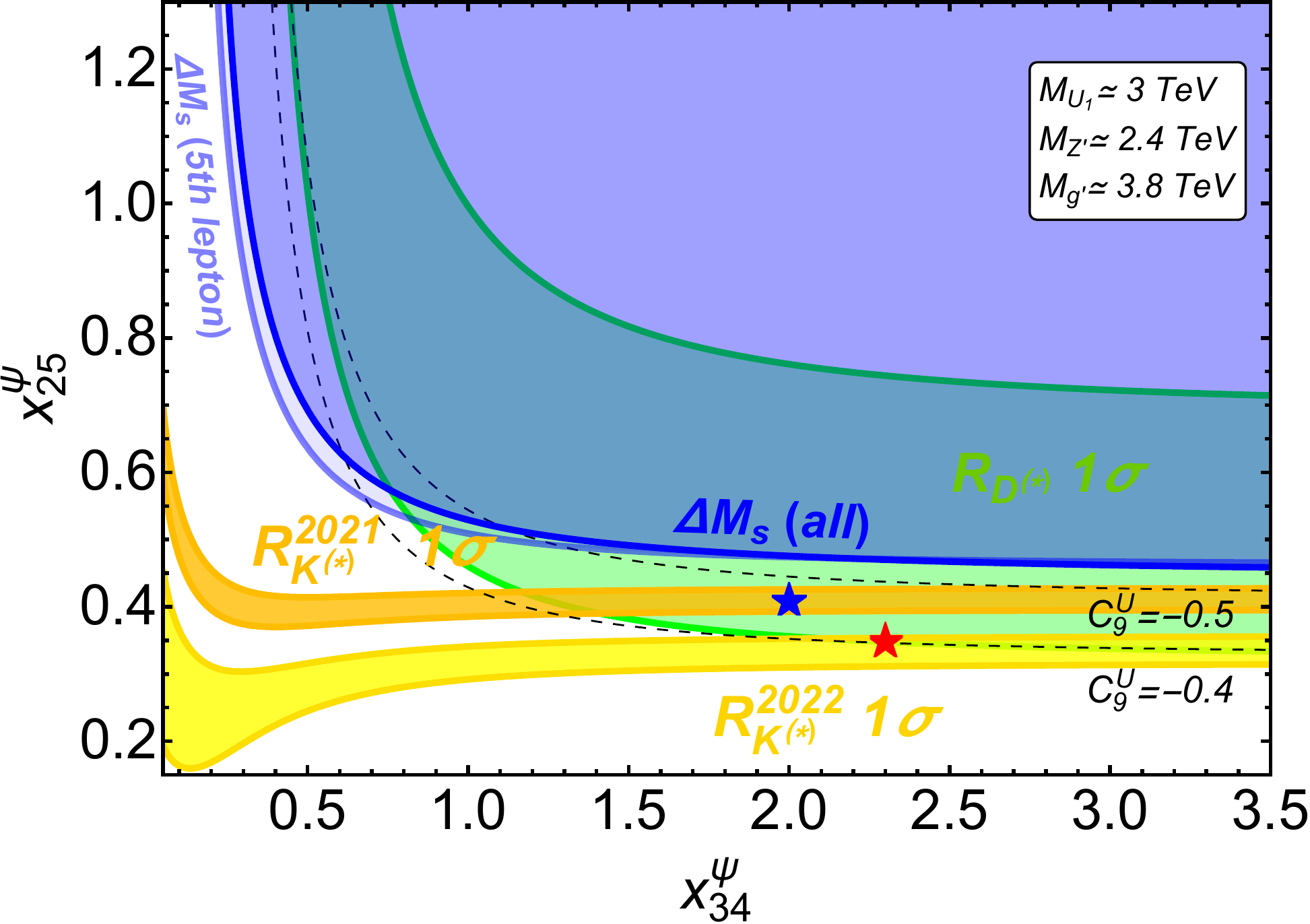}

}$\qquad$\subfloat[\label{fig:Bs_Mixing_ML5}]{\includegraphics[scale=0.37]{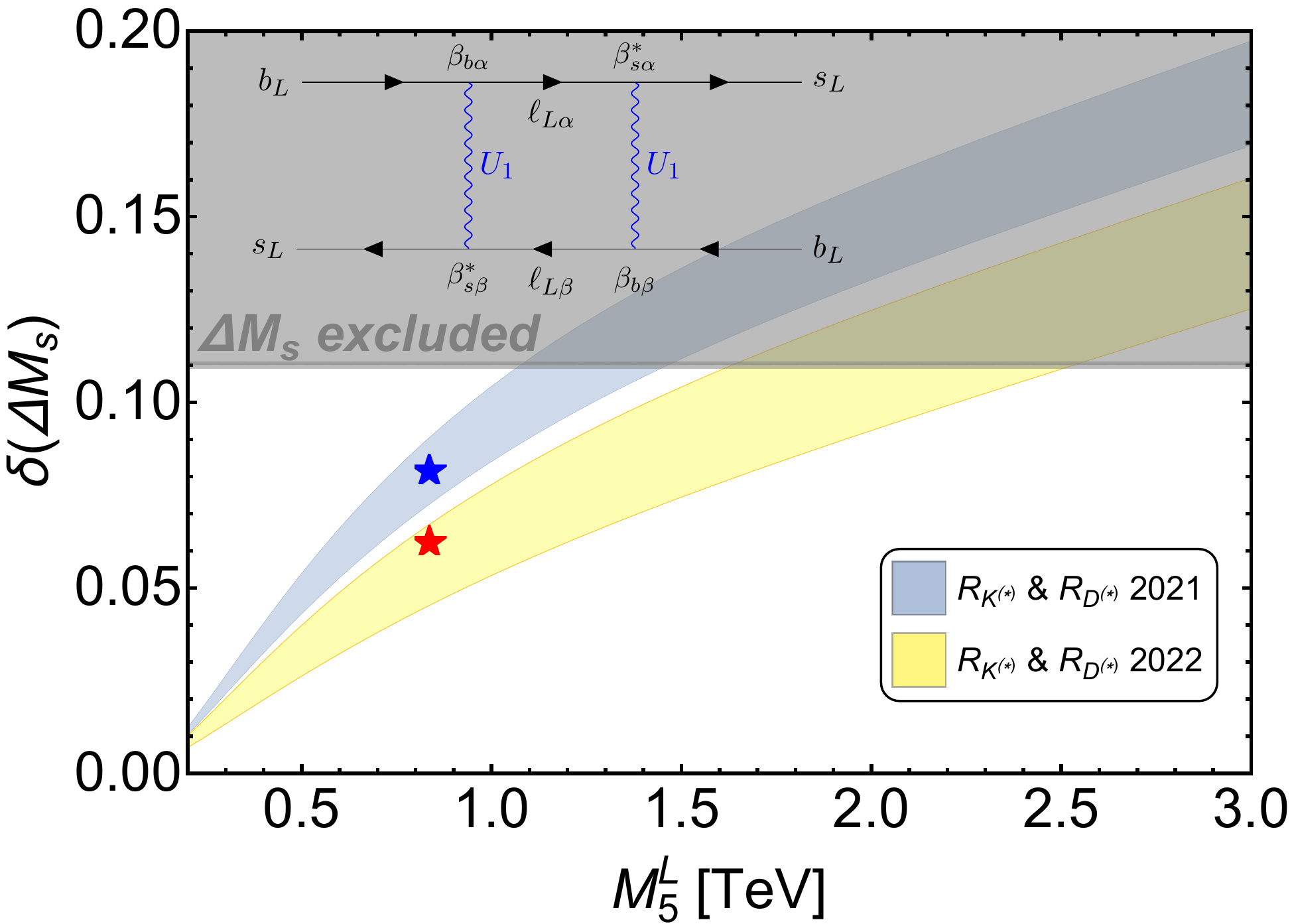}}

\caption{\textbf{\textit{(Left)}} Parameter space in the plane ($x_{34}^{\psi}$,
$x_{25}^{\psi}$) compatible with the LFU ratios. The rest of parameters
are fixed as in the benchmark of \cite{FernandezNavarro:2022gst}
for both panels. The blue region is excluded by the $\Delta M_{s}$
bound, the region excluded only due to the contribution via the 5th
lepton is also shown in lighter blue for comparison. The blue and
red stars show benchmark points for 2021 and 2022 data. \textbf{\textit{(Right)}}\textit{
}$\delta(\Delta M_{s})=\Delta M_{s}^{\mathrm{NP}}/\Delta M_{s}^{\mathrm{SM}}$
as a function of the 5th family lepton vector-like mass term. $x_{25}^{\psi}$
is varied in the range $x_{25}^{\psi}=[0.3,\,0.35]$ ($[0.4,\,0.45]$)
preferred by $R_{K^{(*)}}^{2022}$ ($R_{K^{(*)}}^{2021}$ ), obtaining
the yellow (blue) band. The gray region is excluded by the bound $\delta(\Delta M_{s})<0.11$
\cite{FernandezNavarro:2022gst}. The stars denote benchmark points.\label{fig:B_anomalies_DeltaMs}}
\end{figure}
Fig.~\ref{fig:Bs_Mixing_ParameterSpace} shows that the model is
compatible with both LFU ratios at 1$\sigma$, although the good parameter
space is narrow with 2022 data on $R_{K^{(*)}}$. A large contribution
to $B_{s}-\bar{B}_{s}$ meson mixing, namely to $\Delta M_{s}$, arises
at 1-loop mediated by $U_{1}$ and constrains part of the parameter
space. This loop is dominated by the 5th VL charged lepton, which
is required to be light in order to pass the stringent test of $\Delta M_{s}$,
see Fig.~\ref{fig:Bs_Mixing_ML5}. This way, a light vector-like
lepton is predicted with a mass around 1 TeV, accessible to direct
searches at LHC.

\begin{figure}
\subfloat[\label{fig:tau_3mu}]{\includegraphics[scale=0.37]{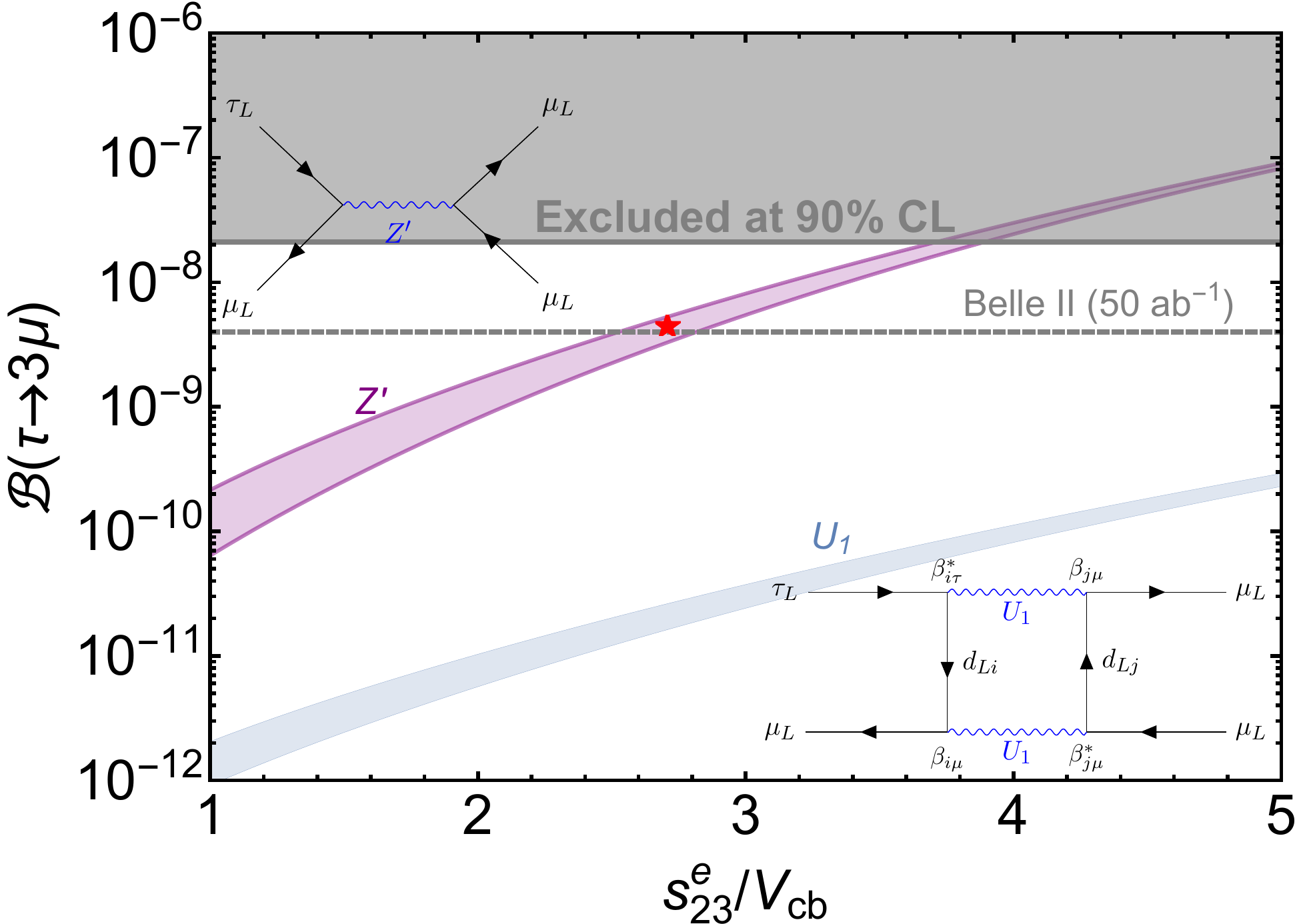}

}$\qquad$\subfloat[\label{fig:tau_mu_gamma}]{\includegraphics[scale=0.365]{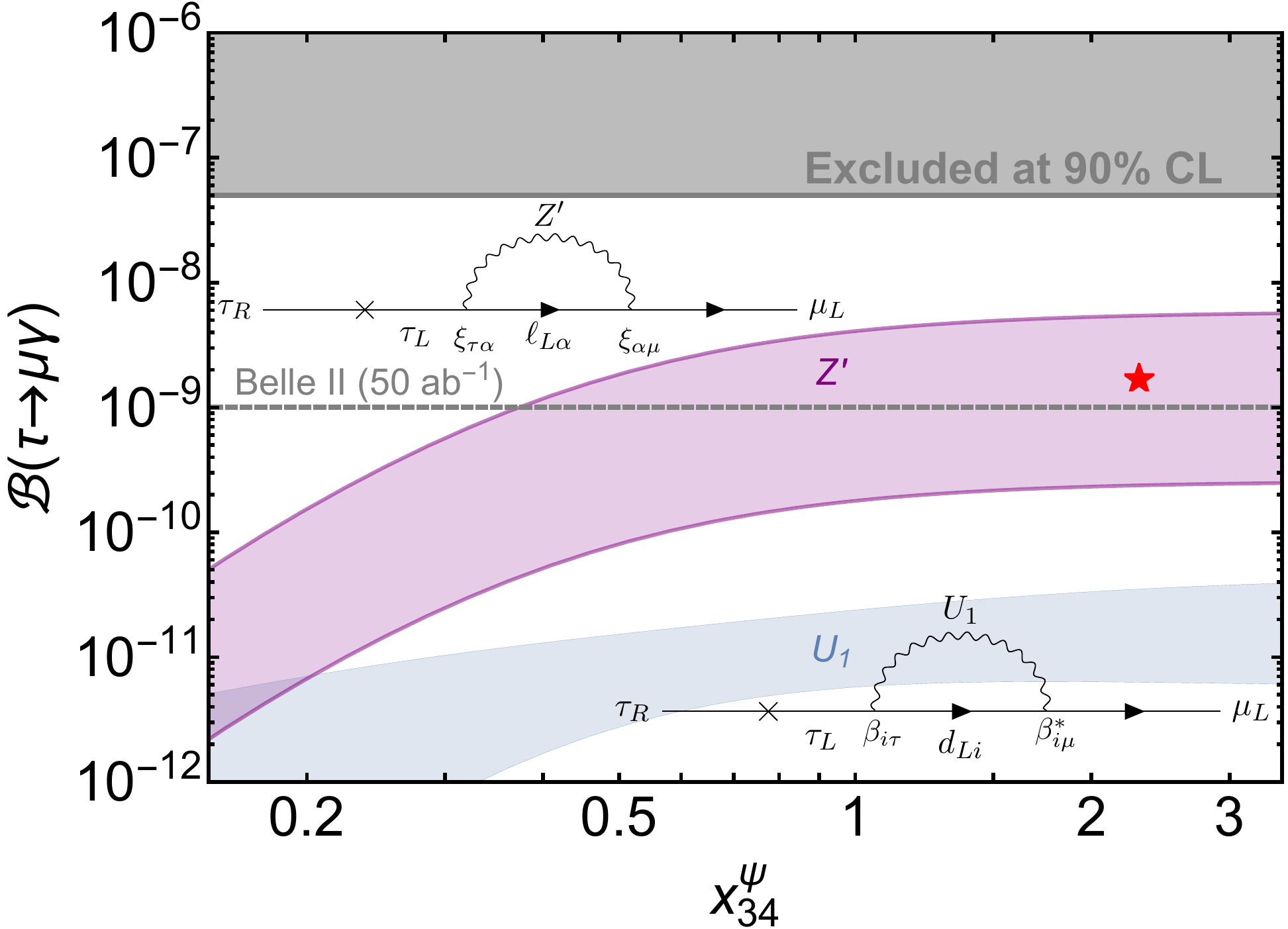}

}

\subfloat[\label{fig:B_K_tautau}]{\includegraphics[scale=0.37]{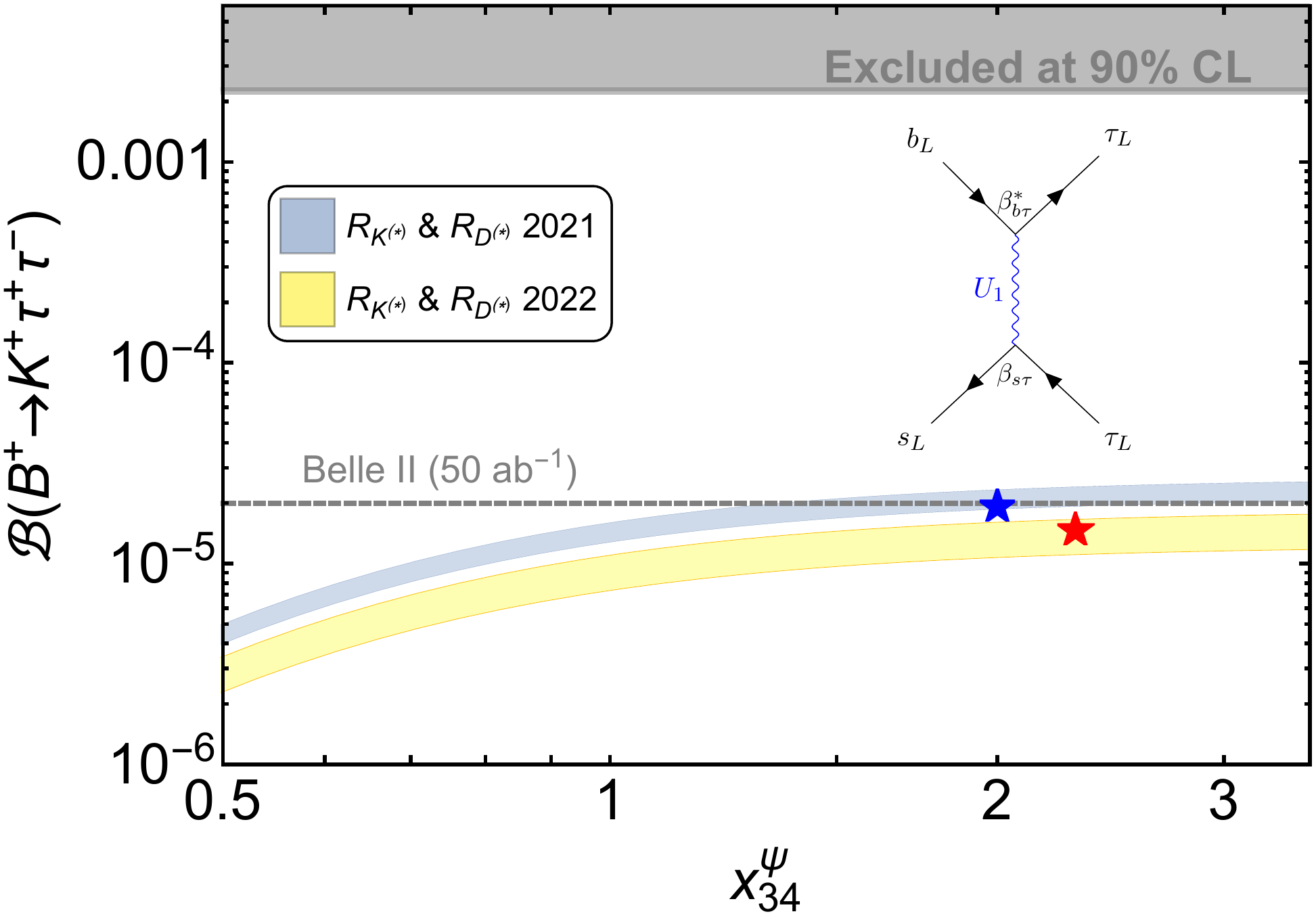}

}$\qquad$\subfloat[\label{fig:B_K_nunu}]{\includegraphics[scale=0.415]{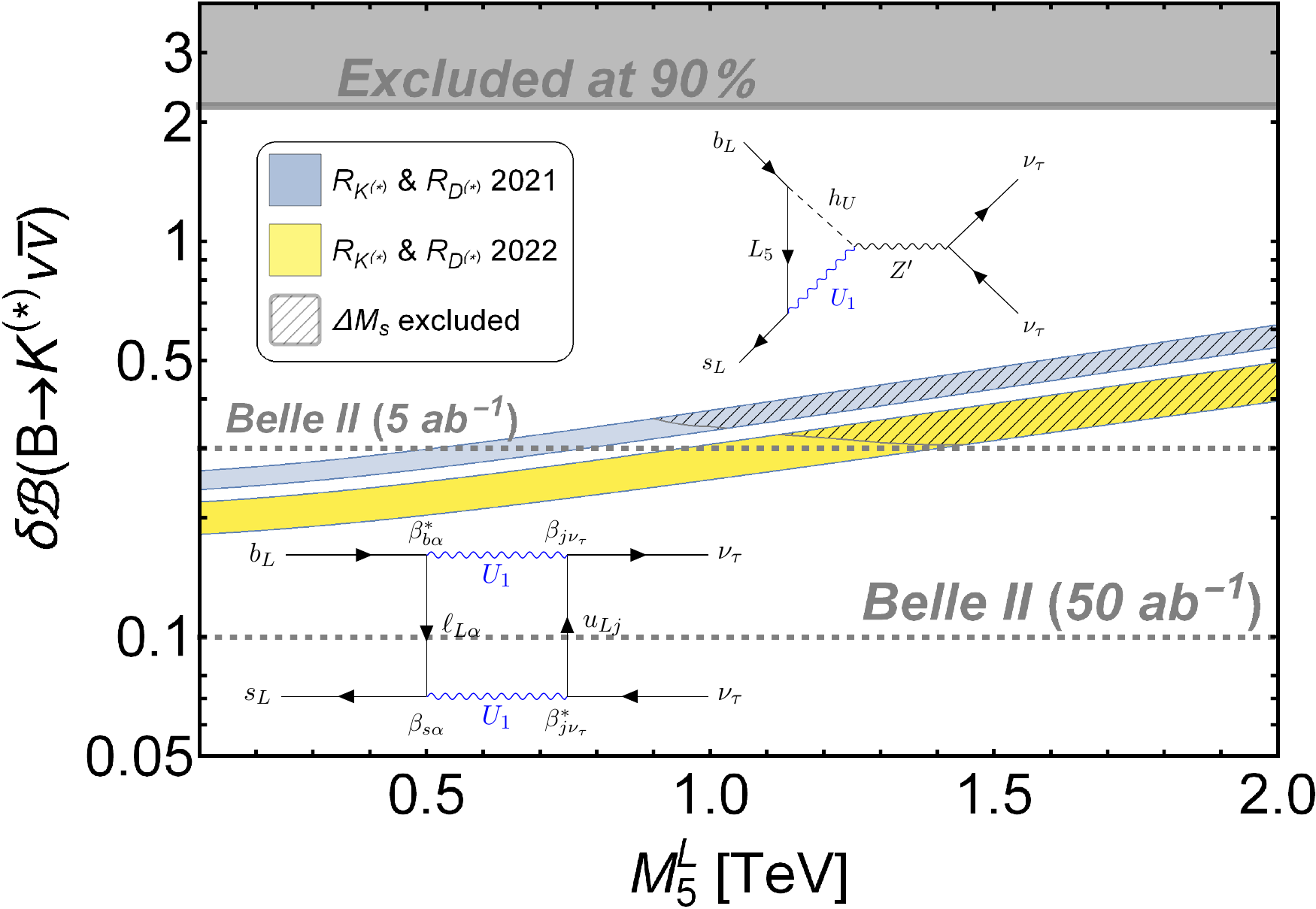}

}

\caption{\textbf{\textit{(Top) }}Branching fractions of LFV $\tau$ decays
as a function of the 2-3 charged lepton mixing sine $s_{23}^{e}$
(for $\tau\rightarrow3\mu$) and $x_{34}^{\psi}$ (for $\tau\rightarrow\mu\gamma$).
The purple region denotes the $Z'$ contribution while the blue region
denotes the $U_{1}$ contribution. For $\tau\rightarrow3\mu$ we have
varied $x_{25}^{\psi}=[0.3,0.35]$ which is compatible with the 2022
LFU ratios, for $\tau\rightarrow\mu\gamma$ we have varied $s_{23}^{e}=[V_{cb},5V_{cb}]$.
\textbf{\textit{(Bottom) }}On the left panel, the branching fraction
of $B\rightarrow K\tau\tau$ as a function of $x_{34}^{\psi}$. On
the right, $\delta(B\rightarrow K\nu\nu)=\mathcal{B}(B\rightarrow K\nu\nu)/\mathcal{B}(B\rightarrow K\nu\nu)_{\mathrm{SM}}-1$
as a function of the 5th family lepton vector-like mass term. In both
panels, $x_{25}^{\psi}$ is varied in the range $x_{25}^{\psi}=[0.3,\,0.35]$
($[0.4,\,0.45]$) preferred by $R_{K^{(*)}}^{2022}$ ($R_{K^{(*)}}^{2021}$
), obtaining the yellow (blue) band. The hatched region is excluded
by the $\Delta M_{s}$ bound, and the stars denote benchmark points.\label{fig:LFV_bsnunu}}
\end{figure}

Signals in $\tau$ LFV decays, as per Figs.~\ref{fig:tau_3mu} and
\ref{fig:tau_mu_gamma}, are directly related to the nature of the
model as a theory of flavour. The model allows for large leptonic
2-3 FCNCs mediated by the $Z'$ boson, and proportional to $\tau-\mu$
mixing arising through a diagram similar to that of Fig.~\ref{fig:2-3_mixing}
but for charged leptons. This $Z'$ contribution is intrinsic to the
twin PS model, and dominates over a further 1-loop contribution mediated
by $U_{1}$ which is common to all 4321 models, hence allowing to
disentangle the twin PS theory from other proposals. A wide range
of the parameter space will be tested in the future by Belle II.

On the other hand, the model predicts large contributions to $b\rightarrow s\tau\tau$
and $b\rightarrow s\nu_{\tau}\nu_{\tau}$ connected to the explanation
of $R_{D^{(*)}}$, as depicted in Figs.~\ref{fig:B_K_tautau} and
\ref{fig:B_K_nunu} . In particular, the full parameter space of the
model will be tested by Belle II via $B\rightarrow K^{(*)}\nu\nu$.

\section{Conclusions}

We have presented a model where the flavour hierarchies are related
to flavour anomalies in $B$-physics. The model features two Pati-Salam
gauge groups broken at different scales, leading to a TeV scale 4321
gauge group. SM-like fermions are all singlets under the TeV scale
$SU(4)$, and hence originally do not couple to the $U_{1}$ vector
leptoquark arising after the breaking of the 4321 symmetry. SM-like
Yukawa couplings are forbidden due to the choice of the scalar sector
along with the twin PS symmetry. Both the effective Yukawa couplings
and the $U_{1}$ couplings for SM fermions arise through mixing with
vector-like fermions transforming under the TeV scale $SU(4)$. This
way, flavour hierarchies and contributions to LFU ratios via $U_{1}$
find a common origin, and are connected via the same physics.

The model predicts a rich phenomenology at low energies, and will
be tested by Belle II via signals in LFV $\tau$ decays and $B\rightarrow K^{(*)}\nu\nu$.
A plethora of new states are predicted at the TeV scale, which could
be directly detected at the LHC, including the heavy $U_{1}$, coloron
and $Z'$. Remarkably, a vector-like lepton is preferred to be around
1 TeV to pass the stringent bounds of $B_{s}-\bar{B}_{s}$ meson mixing.
The model is still compatible with 2022 data on $R_{K^{(*)}}$, although
the window to explain $R_{D^{(*)}}$ at $1\sigma$ is now very narrow.

\providecommand{\href}[2]{#2}\begingroup\raggedright\endgroup

\end{document}